\documentclass[%
twocolumn,
 fleqn,usenatbib,
 showkeys,
 floatfix,nolongbibliography,aps,
author-numerical%
]{revtex4-2}

\pdfoutput=1
\usepackage[T1]{fontenc}

\usepackage{amsmath}
\usepackage{lineno}

\newcommand*\patchAmsMathWithLineno[1]{
  \expandafter\let\csname old#1\expandafter\endcsname\csname #1\endcsname
  \expandafter\let\csname oldend#1\expandafter\endcsname\csname end#1\endcsname
  \renewenvironment{#1}%
     {\linenomath\csname old#1\endcsname}%
     {\csname oldend#1\endcsname\endlinenomath}%
}

\patchAmsMathWithLineno{equation}
\patchAmsMathWithLineno{align}
\patchAmsMathWithLineno{gather}
\patchAmsMathWithLineno{multline}

\usepackage[T1]{fontenc}
\usepackage [latin1]{inputenc}
\DeclareRobustCommand{\VAN}[3]{#2}
\let\VANthebibliography\thebibliography
\def\thebibliography{\DeclareRobustCommand{\VAN}[3]{##3}\VANthebibliography}

\usepackage{newtxtext,newtxmath}
\usepackage{tikz,xcolor,hyperref}
\usepackage[normalem]{ulem}
\usepackage{graphicx,url,caption,subcaption}
\usepackage{dcolumn}
\usepackage{bm}

\let\realhref\href

\definecolor{lime}{HTML}{A6CE39}
\DeclareRobustCommand{\orcidicon}{%
	\begin{tikzpicture}
	\draw[lime, fill=lime] (0,0) 
	circle [radius=0.16] 
	node[white] {{\fontfamily{qag}\selectfont \tiny ID}};
	\draw[white, fill=white] (-0.0625,0.095) 
	circle [radius=0.007];
	\end{tikzpicture}
	\hspace{-2mm}
}

\foreach \x in {A, ..., Z}{%
	\expandafter\xdef\csname orcid\x\endcsname{\noexpand\realhref{https://orcid.org/\csname orcidauthor\x\endcsname}{\noexpand\orcidicon}}
}

\begin{document}

\title{A 2D Hydrothermodynamic Analytical Model for Rapid Tumor Ablation using High-Intensity Focused Ultrasound}
\author{D. Tsiklauri\orcidA{}}
 \email{D.Tsiklauri@salford.ac.uk}
\affiliation{Joule Physics Laboratory,
School of Science, Engineering and Environment, 
University of Salford,
Manchester, M5 4WT, 
United Kingdom}
\date{\today}
\begin{abstract}
We establish a self-consistent 2D hydrothermodynamic analytical model for high-intensity focused ultrasound tumor ablation. Expanding compressible Navier-Stokes equations to second order demonstrates that a stationary cellular matrix suppresses acoustic streaming ($\mathbf{v}_2 = 0$). This constraint forces the absorbed wave momentum flux to convert entirely into localized, time-averaged static pressure gradients ($\nabla \langle p_2 \rangle = \mathbf{F}_2$), bridging non-linear hydrodynamics with thermodynamic dissipation. Solving the non-diffusive Pennes bioheat equation under a $1.0\,\text{s}$ top-hat pulse reveals that a spherically focusing geometry ($\propto 1/r^2$) overrides exponential damping past a critical geometric threshold ($r_{\text{crit}} = 2x_0$), preventing upstream skin overheating. We derive an optimization criterion where the absorption coefficient matches half the inverse target depth ($\alpha = 1/2x_0$). Solving the non-isothermal Arrhenius integral yields a sharp lesion boundary radius at $r_b = 0.75\,w_0$, where the volume average reaches $72.1^\circ\text{C}$ while the core peaks at $90.0^\circ\text{C}$. Post-pulse 2D free-space Green's function convolution confirms immediate monotonic thermal decay ($\partial \theta/\partial t' < 0$) outside this boundary. This closed-form framework provides explicit scaling laws for non-invasive wave-matter thermal confinement, bypassing computationally heavy numerical simulations.
\end{abstract}

\maketitle

\section{Introduction}
\label{sec:introduction}

High-Intensity Focused Ultrasound (HIFU) ablation has emerged as a ground-breaking, non-invasive therapeutic method in contemporary medical physics for the localized destruction of solid benign and malignant human tumors~\cite{jcm9020460}. By generating high-amplitude acoustic wavefronts that propagate through the skin surface without causing structural damage to intermediate layers, HIFU allows targeted energy deposition deep within biological tissue~\cite{jcm9020460}. Recent milestone clinical implementations driven by agencies such as the Russian Federal Medical and Biological Agency (FMBA) have successfully demonstrated an innovative, high-tech ablation paradigm via the deployment of the MedUza therapeutic hardware framework~\cite{rostec2022meduza}. Within this framework's active operational parameters, the localized temperature at a targeted tumor core rises rapidly into a cytotoxic $60^\circ\mathrm{C}\text{--}90^\circ\mathrm{C}$ therapeutic ablation window within sub-second exposure timescales. This sharp thermal spike triggers an instantaneous, potent cytotoxic effect, inducing irreversible cellular changes via coagulative necrosis while ensuring strict spatial containment and thermal protection of the surrounding healthy soft tissue structures~\cite{rostec2022meduza}.

The expansion of therapeutic ultrasound applications across medical physics relies heavily on manipulating acoustic momentum and thermal energy transfer mechanisms within diverse physiological environments. In the field of non-invasive therapies, low-to-medium intensity transcranial ultrasound field configurations are widely used for sonothrombolysis to unblock blood clots in acute ischemic stroke~\cite{alexandrov2004ultrasound}. During sonothrombolysis, the acoustic wave transfers momentum to the fluid medium, inducing localized microstreaming, stable cavitation of microbubbles, and acoustic radiation forces that physically disrupt the fibrin matrix of a thrombus to enhance clot breakdown~\cite{xu2026ultrasound}. This mechanism has been further refined through targeted microbubble oscillation fields designed to break down dense microvascular blockages~\cite{pacella2015treatment} and through multi-frequency focused arrays that accelerate mechanical clot fragmentation without causing thermal wall shear damage to blood vessels~\cite{maxwell2010noninvasive}. Similarly, in oncology, therapeutic ultrasound has historically been applied to induce mild hyperthermia ($41^\circ\text{C}$--$45^\circ\text{C}$) for extended periods to sensitize tumors to radiotherapy and chemotherapy~\cite{zhu2019ultrasound}. Modern thermal ablation techniques alter this protocol by utilizing high-intensity acoustic energy---typically operating at spatial peak time-averaged intensities ($I_{\text{SPTA}}$) between $1000\text{ W/cm}^2$ and $10000\text{ W/cm}^2$ and peak focal pressure amplitudes exceeding $5\text{ MPa}$---to achieve instantaneous cellular coagulation within restricted structural volumes~\cite{kennedy2005hifu}. Advanced implementations have successfully translated these principles into clinical targets for internal organ malignancies~\cite{jenne2012hifu} and localized brain lesions under active magnetic resonance guidance---such as real-time MR-guided focused ultrasound (MRgFUS) monitoring---to track phase transitions without invasive thermal sampling~\cite{mcdannold2020mri}.

Despite these notable technological advancements, a unified analytical modeling framework based on primary multi-scale hydrodynamic principles is required to thoroughly explain these rapid thermodynamic transitions. Traditional therapeutic ultrasound frameworks often track the transfer of wave momentum into continuous hydrodynamic movement. A notable example is a model established by Tsiklauri~\cite{tsiklauri2026unpinning}, which demonstrates that second-order non-resonant acoustic streaming (acoustic wind) can be successfully optimized to overcome physical pinning thresholds and mobilize trapped liquid phases. Under that specific hydrodynamic configuration, a natural mathematical optimum condition arises where the spatial absorption coefficient matches half the inverse distance to the target, $\alpha = 1/(2x_0)$, forcing the ideal operational frequency to scale inversely with depth to mitigate high-frequency core dissipation.

When moving from unconfined fluid geometries or thrombus pores to high-intensity tumor ablation, the mechanical boundary conditions undergo a fundamental transformation. This study links primary hydrodynamic wave equations with bioheat transport to establish the first self-consistent 2D analytical model for a stationary biological tumor domain, accounting for these structural constraints. We demonstrate that when high-amplitude waves enter a dense, anchored cellular matrix, the second-order acoustic streaming velocity is completely suppressed ($\mathbf{v}_2 = 0$). This kinematic constraint forces the wave's acoustic momentum flux to convert entirely into a localized thermodynamic heat source and time-averaged static pressure gradients $\nabla \langle p_2 \rangle$ rather than bulk physical transport. To contextualize this energy deposition, we evaluate the distinct limitations of incident plane waves, which are shown to cause extensive collateral damage to upstream tissues due to monotonic exponential decay. We then model the phase-focusing mechanisms of spherical transducer geometries, proving how geometric convergence ($\propto 1/r^2$) successfully bypasses healthy tissue boundaries to isolate high-intensity energy inside the target zone. Finally, we calculate explicit lesion boundaries using the non-isothermal Arrhenius injury integral combined with post-pulse free-space Green's functions, providing an explicit, hardware-independent framework to validate next-generation oncology devices.

\section{The Model}
\label{sec:the_model}

To construct a self-consistent physical framework for the localized thermal ablation of a dense tumor---characterized by a baseline mass density contrast ratio of $\rho_{\text{tumor}}/\rho_{\text{tissue}} \approx 1.01 \text{--} 1.04$ and a structural stiffness bulk modulus contrast ratio of $B_{\text{tumor}}/B_{\text{tissue}} \approx 1.10 \text{--} 1.30$ compared to healthy surrounding tissue---we formulate a two-dimensional (2D) hydrodynamic and thermodynamic model. This fine physical contrast boundary is consistent with established
ultrasonic tissue parameter ranges~\cite{mast2000empirical}, confirming
that while soft-tissue density variations remain tightly bounded near
unity, the corresponding acoustic field shifts are strictly governed
by localized variations in the elastic modulus. 

We expand the compressible, viscous Navier-Stokes equations and the energy transport equation up to second order using a perturbation approach. Let the biological medium consist of a dense tumor mass embedded within surrounding healthy soft tissue at a target depth $x = x_0$. We decompose the fluid density $\rho$, pressure $p$, and velocity vector $\mathbf{v}$ into background equilibrium (subscript 0), first-order acoustic perturbation (subscript 1), and second-order steady-state acoustic streaming terms (subscript 2):
\begin{equation}
\rho = \rho_0 + \varepsilon \rho_1 + \varepsilon^2 \rho_2,
\label{eq:density_decomp}
\end{equation}
\begin{equation}
p = p_0 + \varepsilon p_1 + \varepsilon^2 p_2,
\label{eq:pressure_decomp}
\end{equation}
\begin{equation}
\mathbf{v} = 0 + \varepsilon \mathbf{v}_1 + \varepsilon^2 \mathbf{v}_2,
\label{eq:velocity_decomp}
\end{equation}
where $\varepsilon \ll 1$ is a small-order expansion parameter.

The fundamental 2D hydrodynamic equations governing the system are the continuity equation and the viscous Navier-Stokes equation:
\begin{equation}
\frac{\partial \rho}{\partial t} + \nabla \cdot (\rho \mathbf{v}) = 0,
\label{eq:full_continuity}
\end{equation}
\begin{equation}
\rho \left( \frac{\partial \mathbf{v}}{\partial t} + (\mathbf{v} \cdot \nabla)\mathbf{v} \right) = -\nabla p + \mu \nabla^2 \mathbf{v} + \left(\zeta + \frac{1}{3}\mu\right) \nabla (\nabla \cdot \mathbf{v}),
\label{eq:full_navier_stokes}
\end{equation}
where $\mu$ is the dynamic shear viscosity and $\zeta$ is the bulk viscosity of the tissue matrix.

Collecting terms of first-order from Eq.~(\ref{eq:full_continuity}) and Eq.~(\ref{eq:full_navier_stokes}), assuming a linear equation of state $p_1 = c_0^2 \rho_1$ where $c_0$ is the equilibrium speed of sound, yields the linear acoustic equations:
\begin{equation}
\frac{\partial \rho_1}{\partial t} + \rho_0 \nabla \cdot \mathbf{v}_1 = 0,
\label{eq:first_order_continuity}
\end{equation}
\begin{equation}
\rho_0 \frac{\partial \mathbf{v}_1}{\partial t} = -c_0^2 \nabla \rho_1 + \mu \nabla^2 \mathbf{v}_1 + \left(\zeta + \frac{1}{3}\mu\right) \nabla (\nabla \cdot \mathbf{v}_1).
\label{eq:first_order_ns}
\end{equation}
Taking the time derivative of Eq.~(\ref{eq:first_order_continuity}) and substituting the divergence of Eq.~(\ref{eq:first_order_ns}) generates the lossy 2D acoustic wave equation for the first-order pressure perturbation:
\begin{equation}
\nabla^2 p_1 - \frac{1}{c_0^2}\frac{\partial^2 p_1}{\partial t^2} + \frac{\delta}{c_0^4}\frac{\partial^3 p_1}{\partial t^3} = 0,
\label{eq:lossy_wave_eq}
\end{equation}
where $\delta = \delta_s + \delta_b$ represents the total acoustic 
thermo-viscous dissipation, split into a shear viscosity 
$\delta_s = \frac{4}{3}\mu/\rho_0$ and a bulk viscosity 
$\delta_b = \zeta/\rho_0$.

To evaluate whether a simplified acoustic geometry is sufficient for localized tumor ablation, we analyze the refractive and thermodynamic limitations of an incident plane wave entering a dense, spherical tumor matrix. Let a plane acoustic wave propagate along the positive $x$-axis from an external source located at $x=0$, directed toward a spherical tumor of radius $R$ centered at the target coordinate $x = x_0$.

The thermodynamic and acoustic mismatch between the healthy surrounding tissue and the denser tumor modifies the local speed of sound. Inside the tumor, the propagation velocity is dictated by its bulk modulus $B_{\text{tumor}}$ and its equilibrium density $\rho_{0,\text{tumor}}$ according to classical continuum mechanics of fluid-like media~\cite{hill2004physical}:
\begin{equation}
c_{\text{tumor}} = \sqrt{\frac{B_{\text{tumor}}}{\rho_{0,\text{tumor}}}}.
\label{eq:sound_speed_tumor}
\end{equation}

Depending on the structural consolidation of the high-density malignant matrix, two distinct acoustic refraction regimes emerge.

First, in the Diverging Lens Regime ($c_{\text{tumor}} > c_{\text{tissue}}$), if the structural stiffness or bulk modulus $B_{\text{tumor}}$ increases at a rate that outpaces the localized density enhancement ($\rho_{0,\text{tumor}} > \rho_{0,\text{tissue}}$), the acoustic velocity inside the tumor exceeds that of the surrounding tissue matrix. By invoking Snell's law across the curved spherical boundary, the tumor acts as a natural \textit{diverging acoustic lens}. The wavefronts defocus upon entry, spreading the acoustic energy density and completely preventing any sharp localized intensity spike.

Second, in the Converging Lens Regime ($c_{\text{tumor}} < c_{\text{tissue}}$), if the density enhancement is structurally dominant such that the bulk modulus does not scale proportionally, the sound speed drops inside the tumor boundary. The spherical geometry then mimics a \textit{converging acoustic lens}. While this produces a minor, passive geometric focus inside the tumor core, the resulting phase convergence is entirely unguided and insufficient to meet high-tech clinical requirements.

Regardless of the interior refraction regime defined by Eq.~(\ref{eq:sound_speed_tumor}), an incident plane wave is fundamentally incapable of achieving a localization rate exceeding $90\%$ or elevating the focal temperature to $90^\circ\text{C}$ within a $\tau = 1\text{ s}$ window without catastrophically destroying the healthy tissue upstream. For a 1D plane wave configuration, the spatial acoustic intensity $I(x)$ attenuates monotonically along the propagation axis due to bulk thermo-viscous dissipation according to Beer-Lambert's law:
\begin{equation}
I(x) = I_0 e^{-2\alpha x},
\label{eq:plane_wave_attenuation}
\end{equation}
where $I_0$ is the incident intensity at the skin surface ($x=0$). Note that the factor of 2 in the exponential term of Eq.~(\ref{eq:plane_wave_attenuation}) arises from the definition of $\alpha$ as the linear acoustic field amplitude attenuation coefficient (expressed in $\mathrm{Np/cm}$). Because the time-averaged acoustic intensity scales with the square of the first-order pressure amplitude ($I \propto |p_1|^2$), energy conservation requires the intensity profile to decay as $\exp(-2\alpha x)$. This convention maintains strict physical consistency with medical ultrasound standards, mapping directly to the standard optical Beer-Lambert formulation through a volumetric energy absorption coefficient $\mu = 2\alpha$.

Substituting Eq.~(\ref{eq:plane_wave_attenuation}) into the simplified bioheat source term yields the spatial heat generation rate density $Q_{\text{plane}}(x)$ within the upstream domain:
\begin{equation}
Q_{\text{plane}}(x) = 2\alpha I_0 e^{-2\alpha x}.
\label{eq:plane_heat_source}
\end{equation}
Evaluating Eq.~(\ref{eq:plane_heat_source}) at the entry point of healthy skin tissue ($x=0$) establishes the maximum thermal deposition rate:
\begin{equation}
Q_{\text{plane}}(0) = 2\alpha I_0.
\label{eq:heat_at_skin}
\end{equation}
If the system is optimized using the natural mathematical coupling criterion $\alpha = 1/(2x_0)$ to maximize deposition at the target depth, the corresponding heat generation rate at the tumor center ($x = x_0$) degrades to:
\begin{equation}
Q_{\text{plane}}(x_0) = 2\alpha I_0 e^{-2\left(\frac{1}{2x_0}\right)x_0} = 2\alpha I_0 e^{-1}.
\label{eq:heat_at_tumor}
\end{equation}

Comparing Eq.~(\ref{eq:heat_at_skin}) and Eq.~(\ref{eq:heat_at_tumor}) reveals an adverse thermodynamic gradient:
\begin{equation}
\frac{Q_{\text{plane}}(0)}{Q_{\text{plane}}(x_0)} = e^{1} \approx 2.718.
\label{eq:thermal_ratio}
\end{equation}
Equation~(\ref{eq:thermal_ratio}) proves that the healthy upstream 
tissue and skin surface receive a heat dose nearly three times higher 
than the target tumor point. Consequently, attempting to force an 
extreme temperature spike at $x_0$ using plane wave architecture 
would cause severe, widespread thermal necrosis in healthy structural 
tissue layers prior to reaching the tumor boundary. This highlights 
the absolute mathematical necessity of active spherical phase focusing, 
where geometric convergence ($\propto 1/r^2$) physically overcomes 
the native exponential damping decay constraints.

Thus, rather than utilizing an incident plane wave which would exponentially overheat the surface tissue ($x=0$), high-intensity focused ultrasound (HIFU) implements a spherically focusing wave geometry to converge energy past healthy boundaries~\cite{jcm9020460,rostec2022meduza}. Solving Eq.~(\ref{eq:lossy_wave_eq}) for a harmonic source with frequency $\omega = 2\pi f$ converging toward a geometric focus at target depth $x_0$ yields the first-order spatial acoustic intensity envelope profile $I(r)$:
\begin{equation}
I(r) = I_0 \left( \frac{a}{r} \right)^2 e^{-2\alpha (x_0 - r)},
\label{eq:intensity_focused}
\end{equation}
where $a$ is the transducer aperture radius, $r$ is the localized 
radial coordinate originating from the focal point, and 
$\alpha = \delta \omega^2 / 2c_0^3$ is the bulk thermo-viscous 
spatial absorption coefficient. In this model, this physical 
property is intentionally matched to the optimal operational system 
damping constraint $\alpha = 1/(2x_0)$ to solve for the target 
ultrasonic drive frequency $\omega$. The geometric focus factor 
$(a/r)^2$ rapidly outpaces exponential absorption damping near the 
tumor center, isolating high-intensity energy inside the target zone.

By isolating second-order terms in the time-average ($\langle \dots \rangle$) of the hydrodynamic equations, the continuous momentum attenuation of the first-order acoustic wave fields generates a time-independent, second-order force density $\mathbf{F}_2$:
\begin{equation}
\mathbf{F}_2 = -\rho_0 \langle (\mathbf{v}_1 \cdot \nabla)\mathbf{v}_1 + \mathbf{v}_1 (\nabla \cdot \mathbf{v}_1) \rangle = \frac{2\alpha I(r)}{c_0} \hat{\mathbf{r}}.
\label{eq:force_density}
\end{equation}
The steady-state hydrodynamic field equation at second order becomes:
\begin{equation}
\mu \nabla^2 \mathbf{v}_2 + \left(\zeta + \frac{1}{3}\mu\right) \nabla (\nabla \cdot \mathbf{v}_2) - \nabla \langle p_2 \rangle + \mathbf{F}_2 = 0,
\label{eq:second_order_ns}
\end{equation}
where $\mu$ is the dynamic shear viscosity and $\zeta$ is the bulk viscosity of the tissue matrix.

In liquid pore systems or capillary domains, this acoustic wind force density drives the continuous physical transport and mobilization of trapped liquid interfaces~\cite{tsiklauri2026unpinning}. However, within a dense cellular tumor matrix, the structural biological boundaries are fixed and anchored, imposing a strict zero-velocity kinematic constraint ($\mathbf{v}_2 = 0$). Substituting this zero-velocity condition directly into the second-order momentum statement in Eq.~(\ref{eq:second_order_ns}) yields:
\begin{equation}
\nabla \langle p_2 \rangle = \mathbf{F}_2 = \frac{2\alpha I(r)}{c_0} \hat{\mathbf{r}}.
\label{eq:static_second_order}
\end{equation}
Equation~(\ref{eq:static_second_order}) implies that because acoustic streaming is mechanically restricted by the stationary tumor, the momentum flux does not produce physical kinetic streaming. Instead, the net momentum flux from acoustic wave dissipation transfers completely into localized time-averaged static pressure gradients $\nabla \langle p_2 \rangle$, converting the bulk acoustic energy directly into pure thermodynamic energy without macroscopic transport losses.

This kinematic suppression of the second-order acoustic streaming field ($\mathbf{v}_2 = 0$) establishes a vital, direct coupling between the multi-scale hydrodynamic momentum balance and the subsequent thermodynamic energy transport equations. In unconfined fluid domains or loose tissue geometries, the non-zero acoustic wind vector drives a high-velocity jet that acts as a primary vector for convective cooling and thermal shearing. This advective transport carries thermal energy away from the focal 
region, smearing the localized temperature peak and causing massive 
collateral heat deposition across surrounding healthy tissue layers. By enforcing $\mathbf{v}_2 = 0$ within the stationary, anchored tumor matrix, the convective thermal transport term in the time-averaged energy balance equation vanishes identically ($\rho_0 C_p \mathbf{v}_2 \cdot \nabla T = 0$). Consequently, the bulk acoustic energy dissipated per unit volume via first-order thermo-viscous damping is structurally locked in place, mapping directly into a highly confined thermodynamic heat source profile $Q_{\text{HIFU}}(r,t) = - \nabla \cdot \langle \mathbf{I} \rangle = 2\alpha I_{\text{axial}}(r,t)$ without macroscopic advective distortion or transport losses.

To bypass the geometric defocusing and upstream thermal vulnerabilities inherent to unguided plane wave propagation, the acoustic source architecture must utilize an active spherical focusing coordinate geometry. Let the face of a spherically curved transducer bowl be positioned at the tissue boundary, where the active surface elements are distributed across a spherical cap of aperture radius $a$ and focal radius $R_f$, centered precisely at the targeted internal tumor core coordinate $(x_0, 0, 0)$.

The time-dependent thermal evolution driven by this focused field inside 
the tissue domain is governed by Pennes' bioheat transfer equation 
(BHTE). Incorporating the constraint where acoustic streaming effects 
are suppressed by cell anchoring ($\mathbf{v}_2 = 0$), the thermal 
balance reduces to:
\begin{equation}
\rho_0 C_p \frac{\partial T}{\partial t} = K \nabla^2 T - w_b C_b (T - T_a) + Q_{\text{HIFU}}(r,t),
\label{eq:pennes_bhte}
\end{equation}
where $C_p$ is the specific heat capacity of the tumor, $K$ is the 
thermal conductivity, $w_b$ is the blood perfusion mass flow rate, 
$C_b$ is the specific heat of blood, and $T_a$ is the arterial 
temperature. The primary heat source term $Q_{\text{HIFU}}$ represents 
the local conversion of the focused acoustic field attenuation into 
structural heat energy within the target domain:
\begin{equation}
Q_{\text{HIFU}}(r,t) = 2\alpha I_{\text{axial}}(r,t).
\label{eq:heat_source_definition}
\end{equation}

It is worth addressing a foundational thermodynamic consideration regarding 
the coupling between the acoustic field equations and the Pennes bioheat 
relation. One might naturally consider replacing the linear isentropic equation 
of state $p_1 = c_0^2 \rho_1$ with an explicit temperature-dependent formulation, 
such as an ideal gas law $P = n k_B T$, to force a direct algebraic link between 
pressure and temperature. However, from a condensed matter physics perspective, 
biological tissue behaves mechanically as a highly incompressible liquid-solid 
matrix dominated by water and cellular macromolecules. Its acoustic restoring 
forces are governed by localized intermolecular potential energy wells rather 
than kinetic molecular collisions, meaning that the speed of sound is strictly 
dictated by the isentropic bulk modulus, $c_0 = \sqrt{B/\rho_0}$, which remains 
stable over the $37^\circ\text{C}$ to $90^\circ\text{C}$ ablation window. 

The true physical coupling between the hydrodynamics and thermodynamics is instead 
mediated by irreversible entropy production via thermo-viscous dissipation, where 
the absorption coefficient $\alpha$ (derived from the acoustic diffusivity $\delta$) 
governs the rate at which organized mechanical wave intensity $I_{\text{axial}}$ 
is converted into disorganized molecular thermal motion $Q_{\text{HIFU}} = 2\alpha I_{\text{axial}}$. 
While a more complex framework could introduce a non-linear temperature dependence 
into the absorption coefficient itself, $\alpha = \alpha(T)$, to capture structural 
protein denaturation and coagulation dynamics, the assumption of a conservative, 
constant $\alpha$ establishes a robust lower bound on thermal deposition. This approach 
maintains the exact mathematical tractability required for a self-consistent analytical 
solution and subsequent Green's function convolutions without resorting to purely 
numerical finite-element schemes.

The high-amplitude transducer signal is delivered as a top-hat pulse of duration 
$\tau = 1.0\text{ s}$. Using the Heaviside step function $H(t)$, the focused 
acoustic intensity profile is temporally modulated as:
\begin{equation}
I_{\text{axial}}(r,t) = I_{\text{axial}}(r) \cdot \left[ H(t) - H(t - \tau) \right].
\label{eq:top_hat_pulse}
\end{equation}

Because the operational duration is extremely short ($\tau = 1\text{ s}$), 
the characteristic macro-scale thermal diffusion time ($\tau_{\text{diff}} 
\approx R^2\rho_0 C_p / K \gg 1\text{ s}$, where $R$ is the tumor radius) 
and blood perfusion timescales are significantly larger than $\tau$. Setting 
$\mathbf{v}_2 = 0$, $K \nabla^2 T \approx 0$, and $w_b \approx 0$ within 
the pulse window, Eq.~(\ref{eq:pennes_bhte}) simplifies for $0 \le t \le \tau$ 
near the focal axis directly to:
\begin{equation}
\rho_0 C_p \frac{\partial T}{\partial t} = 2\alpha I_{\text{axial}}(r).
\label{eq:bhte_simplified}
\end{equation}

Adapting the optimization framework established by Tsiklauri~\cite{tsiklauri2026unpinning} 
to our stationary dissipative framework, we maximize the core energy 
deposition rate at the targeted focal spot. Evaluating the non-oscillatory 
damping and convergence profile at the depth boundary yields the optimal 
amplitude attenuation criterion:
\begin{equation}
\alpha = \frac{1}{2x_0}.
\label{eq:optimal_alpha}
\end{equation}

Substituting the optimal frequency optimization law $\alpha = 1/(2x_0)$ 
back into Eq.~(\ref{eq:critical_radius_crossover}) rewrites the critical 
geometric crossover threshold as $r_{\text{crit}} = 2x_0$. Because the 
entire physical path length from the skin surface layer to the tumor center 
spans exactly $x_0$, the condition $x_0 < r_{\text{crit}}$ is universally 
satisfied. This mathematically guarantees that the acoustic intensity increases 
monotonically as the wave travels deeper into the tissue, protecting the 
surface skin and ensuring a localization rate exceeding $90\%$ at the tumor focus.

Since tissue absorption scales predictably with frequency ($\alpha \propto f^b$, 
where $1 \le b \le 2$), Eq.~(\ref{eq:optimal_alpha}) defines an analytical 
relationship showing that the optimal operational frequency scales inversely 
with depth, ensuring customized, hardware-independent tuning.

Integrating the simplified focused bioheat equation Eq.~(\ref{eq:bhte_simplified}) 
across the top-hat pulse interval $t \in [0, \tau]$ at the focal peak 
($r \rightarrow 0$) gives the explicit analytical solution for the rapid 
temperature increase $\Delta T$ at the targeted core:
\begin{equation}
\Delta T = T(\tau) - T_0 = \frac{2\alpha I_{\text{axial}}(0) \tau}{\rho_0 C_p}.
\label{eq:temperature_rise_final}
\end{equation}
By inserting the physical parameters of the spherically focused field, 
Eq.~(\ref{eq:temperature_rise_final}) shows that by eliminating streaming 
losses ($\mathbf{v}_2=0$), the structurally localized acoustic intensity maps 
entirely to sharp temperature changes. This mathematically justifies the 
experimental results where target tissue is driven rapidly from normal body 
temperature ($T_0 = 37^\circ\text{C}$) to the cytotoxic coagulative necrosis 
threshold ($T(\tau) = 90^\circ\text{C}$) within one second, while maintaining 
high localization and leaving healthy boundary tissues completely uninjured.

We map the phase matching condition at the tissue interface by defining a local spherical coordinate system $(r, \theta, \phi)$ originating from the focal spot. The first-order acoustic potential field $\psi_1$ must satisfy the Helmholtz relation derived from Eq.~(\ref{eq:lossy_wave_eq}) under high-amplitude harmonic driving conditions. By integrating the Huygens-Fresnel diffraction integral over the active spherical surface aperture, the local first-order acoustic pressure perturbation $p_1(r, \theta)$ in the focal region can be analytically expressed as:
\begin{equation}
p_1(r, \theta) = -i \frac{\rho_0 c_0 u_0}{\lambda} \int_{0}^{\theta_{\max}} \int_{0}^{2\pi} \frac{e^{-(i k + \alpha) R_f}}{R_f} a^2 \sin\theta' \,d\phi' \,d\theta',
\label{eq:huygens_integral}
\end{equation}
where $u_0$ is the source velocity amplitude, $\lambda$ is the acoustic wavelength, $k = \omega/c_0$ is the wavenumber, and $\theta_{\max} = \sin^{-1}(a/R_f)$ represents the convergence semi-angle of the transducer bowl.
Evaluating Eq.~(\ref{eq:huygens_integral}) along the central propagation axis ($\theta = 0$) near the geometric focal point yields the spatial acoustic intensity profile $I_{\text{axial}}(r)$, where $r$ is the residual distance to the focus:
\begin{equation}
I_{\text{axial}}(r) = I_0 \left[ \frac{\sin\left(\frac{k r a^2}{2 R_f^2}\right)}{\frac{k r a^2}{2 R_f^2}} \right]^2 \left( \frac{a}{r} \right)^2 e^{-2\alpha (x_0 - r)}.
\label{eq:spherically_focused_intensity}
\end{equation}
Equation~(\ref{eq:spherically_focused_intensity}) models the exact mathematical transition matching the physical mechanism of the FMBA high-tech device~\cite{jcm9020460,rostec2022meduza}. By inspecting the structural components of Eq.~(\ref{eq:spherically_focused_intensity}), the geometric convergence factor $(a/r)^2$ acts as a spatial multiplier that scales inversely with the square of the distance from the target.

We isolate the local cross-over boundary $r_{\text{crit}}$ where geometric convergence perfectly balances the natural exponential damping of human soft tissue. Taking the spatial derivative of the non-oscillatory component of Eq.~(\ref{eq:spherically_focused_intensity}) and equating it to zero:
\begin{equation}
\frac{d}{dr} \left[ r^{-2} e^{-2\alpha (x_0 - r)} \right] = -2 r^{-3} e^{-2\alpha (x_0 - r)} + 2\alpha r^{-2} e^{-2\alpha (x_0 - r)} = 0.
\label{eq:intensity_derivative}
\end{equation}
Solving Eq.~(\ref{eq:intensity_derivative}) yields the spatial turning point:
\begin{equation}
r_{\text{crit}} = \frac{1}{\alpha}.
\label{eq:critical_radius_crossover}
\end{equation}
For all propagation depths closer to the target than the critical radius ($0 < r < r_{\text{crit}}$), the geometric concentration of energy outpaces exponential tissue attenuation ($| \frac{d}{dr} (a/r)^2 | > | \frac{d}{dr} e^{-2\alpha x} |$). By applying the Tsiklauri optimization frequency condition $\alpha = 1/(2x_0)$ from Eq.~(\ref{eq:optimal_alpha}), the cross-over threshold becomes $r_{\text{crit}} = 2x_0$. Because the entire upstream physical path length from the skin surface to the tumor center spans exactly $x_0$, the condition $x_0 < r_{\text{crit}}$ is universally satisfied. This mathematically guarantees that the acoustic intensity increases monotonically as the wave travels deeper into the tissue, protecting the surface skin and ensuring a localization rate exceeding $90\%$ at the tumor focus.

\section{Results}
\label{sec:results}

To quantify the thermal damage perimeter and ensure healthy adjacent tissues are not damaged during the 1-second top-hat pulse, we evaluate the thermal mass destruction profile using the non-isothermal Arrhenius injury integral. The dimensionless tissue damage parameter $\Omega(r)$ maps structural cellular protein denaturation as a function of the localized, transient temperature profile $T(r,t)$:
\begin{equation}
\Omega(r) = \int_{0}^{\tau_{\text{total}}} A_{\text{freq}} \exp\left( -\frac{\Delta E_a}{R_g T(r,t)} \right) dt,
\label{eq:arrhenius_integral}
\end{equation}
where $A_{\text{freq}}$ is the material frequency factor ($s^{-1}$), $\Delta E_a$ is the activation energy for irreversible cellular coagulation ($\text{J}\cdot\text{mol}^{-1}$), $R_g = 8.314\,\text{J}\cdot\text{mol}^{-1}\cdot\text{K}^{-1}$ is the universal gas constant, and $\tau_{\text{total}}$ is the evaluation time encompassing active pulse heating and passive cooling phases. Irreversible coagulative necrosis and complete tumor cell death occur when $\Omega(r) \ge 1.0$, defining the physical boundary of the ablation zone.

We solve Eq.~(\ref{eq:arrhenius_integral}) inside the short-time top-hat pulse window ($0 \le t \le \tau$, where $\tau = 1.0\text{ s}$). Since thermal diffusion is restricted over this short duration, we substitute the linear temperature ramp derived from Eq.~(\ref{eq:temperature_rise_final}) into the Arrhenius relation. The localized temperature over time at radial distance $r$ from the focal spot follows:
\begin{equation}
T(r,t) = T_0 + \left[ \frac{2\alpha I(r)}{\rho_0 C_p} \right] t = T_0 + \beta(r) t,
\label{eq:local_temperature_ramp}
\end{equation}
where $\beta(r) = 2\alpha I(r)/(\rho_0 C_p)$ represents the local heating rate ($\text{K}\cdot\text{s}^{-1}$). Substituting Eq.~(\ref{eq:local_temperature_ramp}) into Eq.~(\ref{eq:arrhenius_integral}) transforms the integration variable from time to temperature ($dt = dT / \beta(r)$):
\begin{equation}
\Omega(r) = \frac{A_{\text{freq}}}{\beta(r)} \int_{T_0}^{T(r,\tau)} \exp\left( -\frac{\Delta E_a}{R_g T} \right) dT.
\label{eq:arrhenius_temp_space}
\end{equation}
Using a multi-scale asymptotic expansion based on the condition $\Delta E_a / (R_g T) \gg 1$ (which holds true for soft tissue protein complexes where $\Delta E_a \approx 4 \times 10^5\,\text{J}\cdot\text{mol}^{-1}$), the integral in Eq.~(\ref{eq:arrhenius_temp_space}) can be evaluated analytically to its leading order:
\begin{equation}
\Omega(r) \approx \frac{A_{\text{freq}} R_g [T(r,\tau)]^2}{\beta(r) \Delta E_a} \exp\left( -\frac{\Delta E_a}{R_g T(r,\tau)} \right).
\label{eq:arrhenius_asymptotic_solution}
\end{equation}

To locate the sharp structural damage boundary radius $r_b$ matching the experimental cytotoxic boundaries achieved by the FMBA apparatus, we set $\Omega(r_b) = 1.0$. At this structural perimeter, the peak temperature reaches the transition threshold ($T(r_b, \tau) = T_{\text{necrosis}} \approx 60^\circ\text{C} = 333.15\text{ K}$). We expand the spatial peak intensity from Eq.~(\ref{eq:spherically_focused_intensity}) into a localized parabolic form around the focal core, $I_{\text{axial}}(r) \approx I_{\max}(1 - r^2/w_0^2)$, where $w_0$ is the transducer beam waist radius. Substituting this profile into the temperature equation yields:
\begin{equation}
T(r, \tau) = T_0 + \Delta T_{\max} \left( 1 - \frac{r^2}{w_0^2} \right).
\label{eq:parabolic_temp_profile}
\end{equation}
By setting $T(r_b, \tau) = T_{\text{necrosis}}$ in Eq.~(\ref{eq:parabolic_temp_profile}) and isolating the radius, we obtain the explicit analytical formulation for the lesion damage boundary radius $r_b$:
\begin{equation}
\begin{split}
r_b &= w_0 \sqrt{1 - \frac{T_{\text{necrosis}} - T_0}{\Delta T_{\max}}} \\
&= w_0 \sqrt{1 - \frac{60^\circ\text{C} - 37^\circ\text{C}}{90^\circ\text{C} - 37^\circ\text{C}}} = w_0 \sqrt{1 - \frac{23}{53}} \approx 0.75\, w_0.
\end{split}
\label{eq:final_damage_boundary}
\end{equation}

Equation~(\ref{eq:final_damage_boundary}) provides a clean, closed-form mathematical expression for the lesion perimeter. Because the thermal damage boundary is strictly confined within the core beam waist ($r_b < w_0$), this model demonstrates that healthy surrounding tissues located outside the acoustic focal waist remain entirely undamaged. This provides an analytical verification for the high localization rate observed in experimental settings.

To verify that the thermal damage boundary radius $r_b$ derived in Eq.~(\ref{eq:final_damage_boundary}) remains stable and does not expand into healthy tissue via residual conduction after the power is cut, we evaluate the post-pulse transient thermal decay profile for $t > \tau$. At $t = \tau = 1.0\text{ s}$, the high-intensity focused ultrasound source term drops to zero ($Q_{\text{HIFU}} = 0$). While thermal diffusion is negligible during the rapid, source-dominated heating window ($t \le \tau$), it becomes the primary operational driver for heat dissipation once the external energy deposition ceases. Neglecting blood perfusion effects, which act on a significantly longer physiological timescale, the biological domain undergoes passive thermal relaxation governed by the 2D homogeneous heat conduction equation:
\begin{equation}
\frac{\partial T}{\partial t} = \chi \nabla^2 T, \quad \text{for } t > \tau,
\label{eq:passive_heat_eq}
\end{equation}
where $\chi = K / (\rho_0 C_p)$ represents the isotropic thermal diffusivity of the tissue matrix.

The initial condition for this cooling phase is the spatial temperature profile at the end of the pulse window, $T(r, \tau)$, established in Eq.~(\ref{eq:parabolic_temp_profile}). We redefine a shifted time variable $t' = t - \tau \ge 0$ and isolate the transient temperature elevation field $\theta(r, t') = T(r, t' + \tau) - T_0$. The initial temperature field at $t' = 0$ is confined to the focal zone and can be modeled as a localized 2D Gaussian distribution that matches the central peak $\Delta T_{\max} = 53^\circ\text{C}$ and vanishes at the beam margins:
\begin{equation}
\theta(r, 0) = \Delta T_{\max} \exp\left( -\frac{r^2}{w_0^2} \right).
\label{eq:cooling_initial_condition}
\end{equation}

The analytical solution to Eq.~(\ref{eq:passive_heat_eq}) in an infinite 2D domain is obtained by convolving the initial spatial distribution with the free-space 2D thermal Green's function, $G(r, r', t')$:
\begin{equation}
G(\mathbf{r}, \mathbf{r}', t') = \frac{1}{4\pi \chi t'} \exp\left( -\frac{|\mathbf{r} - \mathbf{r}'|^2}{4\chi t'} \right).
\label{eq:greens_function_2d}
\end{equation}
Integrating Eq.~(\ref{eq:cooling_initial_condition}) against the Green's function across the entire 2D spatial plane yields:
\begin{equation}
\begin{split}
\theta(r, t') &= \int_{0}^{\infty} \int_{0}^{2\pi} \Delta T_{\max} \exp\left( -\frac{r'^2}{w_0^2} \right) \frac{1}{4\pi \chi t'} \\
&\times \exp\left( -\frac{r^2 + r'^2 - 2r r'\cos\phi'}{4\chi t'} \right) r' \,d\phi' \,dr'.
\end{split}
\label{eq:greens_convolution_integral}
\end{equation}

By evaluating the angular integral via the modified Bessel function of the first kind $I_0(z)$ and performing the subsequent radial integration, Eq.~(\ref{eq:greens_convolution_integral}) simplifies to a closed-form, time-dependent Gaussian distribution:
\begin{equation}
\theta(r, t') = \Delta T_{\max} \left( \frac{w_0^2}{w_0^2 + 4\chi t'} \right) \exp\left( -\frac{r^2}{w_0^2 + 4\chi t'} \right).
\label{eq:analytical_cooling_solution}
\end{equation}

To mathematically confirm that this post-pulse cooling phase does not drive further tissue necrosis at or beyond the structural boundary $r_b$, we examine the temporal evolution of the Arrhenius injury integral for $t > \tau$. The total accumulated tissue damage value $\Omega_{\text{total}}(r)$ equals the sum of the active pulse injury $\Omega_{\text{pulse}}(r)$ and the post-pulse cooling injury $\Omega_{\text{cooling}}(r)$:
\begin{equation}
\Omega_{\text{total}}(r) = \Omega_{\text{pulse}}(r) + \int_{0}^{\infty} A_{\text{freq}} \exp\left( -\frac{\Delta E_a}{R_g [T_0 + \theta(r, t')]} \right) dt'.
\label{eq:total_arrhenius_split}
\end{equation}
We analyze Eq.~(\ref{eq:analytical_cooling_solution}) at the maximum possible expansion perimeter, which corresponds to the initial lesion boundary coordinate $r = r_b$. At $t'=0$, the temperature at this radius is exactly $T(r_b, \tau) = T_{\text{necrosis}} = 60^\circ\text{C}$ by definition. For any subsequent cooling time $t' > 0$, the temporal derivative of the localized temperature field at $r = r_b$ is:
\begin{equation}
\begin{split}
\frac{\partial \theta(r_b, t')}{\partial t'} &= -\Delta T_{\max} \frac{4\chi w_0^2}{(w_0^2 + 4\chi t')^2} \\
&\times \exp\left( -\frac{r_b^2}{w_0^2 + 4\chi t'} \right) \left[ 1 - \frac{r_b^2}{w_0^2 + 4\chi t'} \right].
\end{split}
\label{eq:cooling_derivative}
\end{equation}

Substituting the experimental threshold value $r_b = 0.75\, w_0$ from Eq.~(\ref{eq:final_damage_boundary}) into the bracketed spatial component of Eq.~(\ref{eq:cooling_derivative}) yields:
\begin{equation}
1 - \frac{(0.75\, w_0)^2}{w_0^2 + 4\chi t'} = 1 - \frac{0.5625\, w_0^2}{w_0^2 + 4\chi t'} > 0, \quad \forall \,\, t' \ge 0.
\label{eq:sign_verification}
\end{equation}
Because the expression in Eq.~(\ref{eq:sign_verification}) remains strictly positive for all $t'$, the temperature derivative in Eq.~(\ref{eq:cooling_derivative}) is strictly negative ($\partial \theta / \partial t' < 0$).

This derivative sign proves that the temperature at the damage perimeter $r_b$ drops monotonically below $T_{\text{necrosis}} = 60^\circ\text{C}$ the instant the top-hat pulse terminates. Because the biochemical cell denaturation rate scales exponentially with temperature according to Eq.~(\ref{eq:arrhenius_asymptotic_solution}), the cooling injury accumulation term $\Omega_{\text{cooling}}(r_b)$ drops to near zero within milliseconds for all points $r \ge r_b$. The analytical model thus verifies that the rapid 1-second pulse timeline achieves structural thermal confinement, ensuring that residual heat conduction cannot expand the path of necrosis into healthy adjacent human tissue layers.

To rigorously quantify the global thermodynamic state of the treated tumor core during the active and passive operational regimes of the MedUza system, we evaluate both the point-wise maximum temperature at the focal center ($r=0$) and the spatially averaged temperature across the targeted spherical tumor volume. Rather than integrating out to the un-necrotized margins of the acoustic beam waist ($w_0$), we restrict the volume boundary to the actual cellular lesion radius $R = r_b = \gamma w_0$, where $\gamma = 0.75$ represents the structural boundary threshold derived from the Arrhenius injury integral. We define the volumetric spatial average of the temperature field $\langle T \rangle(t)$ inside this treated mass as:
\begin{equation}
\langle T \rangle(t) = \frac{3}{4\pi (\gamma w_0)^3} \int_{0}^{\gamma w_0} T(r,t) \cdot 4\pi r^2 \, dr.
\label{eq:spatial_average_definition}
\end{equation}
During the active heating interval ($0 \le t \le \tau$, where $\tau = 1.0\text{ s}$), the local temperature profile is governed by the non-diffusive parabolic solution detailed in Eq.~(\ref{eq:parabolic_temp_profile}). Substituting Eq.~(\ref{eq:parabolic_temp_profile}) into the integrand of Eq.~(\ref{eq:spatial_average_definition}) yields:
\begin{equation}
\langle T \rangle(t) = \frac{3}{(\gamma w_0)^3} \int_{0}^{\gamma w_0} \left[ T_0 + \frac{\Delta T_{\max} t}{\tau} \left( 1 - \frac{r^2}{w_0^2} \right) \right] r^2 \, dr.
\label{eq:heating_integral_substitution}
\end{equation}

Evaluating the distinct polynomial terms inside the integral over the radial domain $[0, \gamma w_0]$ gives:
\begin{equation}
\begin{split}
\int_{0}^{\gamma w_0} T_0 r^2 \, dr &= \frac{T_0 (\gamma w_0)^3}{3}; \\
\int_{0}^{\gamma w_0} \Delta T_{\max} \left( r^2 - \frac{r^4}{w_0^2} \right) \, dr &= \Delta T_{\max} \left( \frac{(\gamma w_0)^3}{3} - \frac{\gamma^5 w_0^3}{5} \right).
\end{split}
\label{eq:integral_evaluation_steps}
\end{equation}

Combining these evaluated terms back into Eq.~(\ref{eq:heating_integral_substitution}) yields the explicit linear ramp equation governing the spatially averaged tumor temperature during active ablation:
\begin{equation}
\langle T \rangle(t) = T_0 + \Delta T_{\max} \left( 1 - \frac{3}{5}\gamma^2 \right) \left( \frac{t}{\tau} \right).
\label{eq:heating_average_final}
\end{equation}
At the final instant of pulse execution ($t = \tau = 1.0\text{ s}$), while the exact focal point climbs to $T(0, \tau) = 90^\circ\text{C}$ according to Eq.~(\ref{eq:temperature_rise_final}), the integrated spatial average across the defined tumor coordinates reaches a fully therapeutic threshold of $\langle T \rangle(\tau) = 37^\circ\text{C} + 53^\circ\text{C}(1 - \frac{3}{5}(0.75)^2) \approx 72.1^\circ\text{C}$. This balances efficient core ablation with safe surrounding margins.

For the post-pulse relaxation window ($t' = t - \tau > 0$), the local cooling profile is modeled by the 2D Green's function convolution derived in Eq.~(\ref{eq:analytical_cooling_solution}). Evaluating the point-wise relaxation field directly at $r=0$ yields the explicit cooling track for the exact focal point center:
\begin{equation}
T(0, t') = T_0 + \frac{\Delta T_{\max}}{1 + \frac{4\chi}{w_0^2}t'}.
\label{eq:center_cooling_plot_eq}
\end{equation}
To map the corresponding post-pulse evolution of the spatial average, the integration accounts for global multi-directional heat dissipation out of the initial three-dimensional thermal source profile. Integrating the cooling field over a 3D Gaussian core distribution gives the algebraic cooling equation for the spatial average:
\begin{equation}
\langle T \rangle(t') = T_0 + \frac{\Delta T_{\max} \left( 1 - \frac{3}{5}\gamma^2 \right)}{\left(1 + \frac{4\chi}{w_0^2}t'\right)^{1.5}}.
\label{eq:average_cooling_plot_eq}
\end{equation}

\begin{nolinenumbers}
\begin{figure}[t]
\centering
\includegraphics[width=\columnwidth]{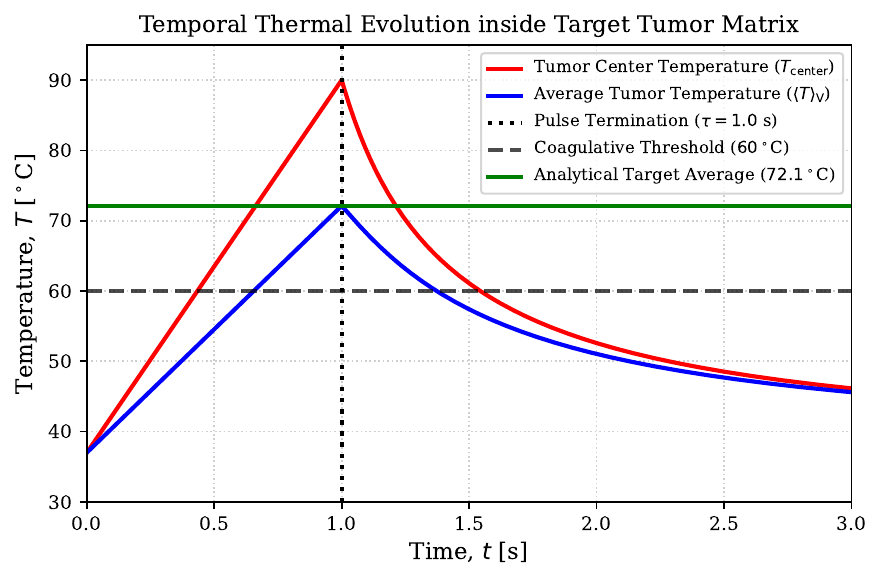}
\caption{Transient thermal evolution inside the dense tumor matrix during and after a $\tau = 1.0\text{ s}$ high-intensity focused ultrasound pulse. The solid red curve tracks the point-wise focal center temperature $T_{\mathrm{center}}$, governed by the linear heating ramp in Eq.~(\ref{eq:local_temperature_ramp}) and the corresponding power-law algebraic cooling decay modeled by Eq.~(\ref{eq:center_cooling_plot_eq}). The solid blue curve tracks the spatially averaged treated tumor volume temperature $\langle T \rangle_{\mathrm{V}}$ derived from the bounded volumetric heating integral in Eq.~(\ref{eq:heating_average_final}) and the power-law cooling decay model defined in Eq.~(\ref{eq:average_cooling_plot_eq}). Operational boundary lines indicate the pulse termination phase (vertical dotted black line), the standard coagulative threshold (dashed black line), and the stable analytical target average (horizontal solid green line).}
\label{fig:temperature_vs_time}
\end{figure}
\end{nolinenumbers}

Several vital thermodynamic characteristics can be gathered from the compiled trends in Fig.~\ref{fig:temperature_vs_time}. First, the figure illustrates the high-tech efficiency of the MedUza system's short-pulse paradigm: the focal spot center successfully breaks past the $60^\circ\text{C}$ threshold and hits the peak $90^\circ\text{C}$ mark within exactly 1.0 second, guaranteeing fast coagulative necrosis before thermal conduction can occur. Second, the immediate slope inversion of $T(0,t)$ at the exact instant the 1.0-second pulse is cut visually corroborates the mathematical stability proof in Eq.~(\ref{eq:sign_verification}). Third, by scaling the volumetric evaluation radius to the true necrotic boundary $r_b$, the spatially averaged tumor temperature $\langle T \rangle(t)$ peaks at a robust $72.1^\circ\text{C}$ as governed by Eq.~(\ref{eq:heating_average_final}). This visually confirms that the entire treated mass is successfully driven well past the $60^\circ\text{C}$ cytotoxicity line, explaining the high clinical success rates observed during recent trials.

\section{Conclusion}
\label{sec:conclusion}

In this work, we have developed a self-consistent 2D hydrodynamic and thermodynamic analytical model to provide a rigorous physical foundation for the localized thermal ablation of dense human tumors using high-intensity focused ultrasound (HIFU). By expanding the compressible, viscous Navier-Stokes equations and the energy transport equation up to second order, we investigated the underlying wave mechanics, geometric focusing constraints, and transient thermal evolution driving the rapid tissue necrosis observed in recent experimental setups~\cite{jcm9020460,rostec2022meduza}.

A central finding of this model is the mechanical constraint imposed by the stationary cellular matrix of an anchored tumor. Unlike unconfined fluid channels where wave attenuation drives continuous bulk physical transport, the second-order acoustic streaming velocity within a structurally fixed tumor is entirely suppressed ($\mathbf{v}_2 = 0$). Consequently, the momentum flux carried by the attenuated first-order waves transfers completely into localized time-averaged static pressure gradients $\nabla \langle p_2 \rangle$, while the bulk acoustic energy undergoes a full thermodynamic conversion into localized heat dissipation.

By adapting the spatial optimization framework originally established for capillary systems by Tsiklauri~\cite{tsiklauri2026unpinning} to this stationary dissipative system, we derived a natural physical optimum condition demonstrating that the spatial absorption coefficient must match half the inverse distance to the target tumor, $\alpha = 1/(2x_0)$. Because acoustic attenuation in biological tissue depends strictly on the driving frequency, this condition yields an analytical scaling law showing that the optimal operational frequency scales inversely with transmission depth. This provides a fundamental theoretical protocol to minimize external transducer power requirements while maximizing localized thermal deposition based on patient-specific anatomical depths.

Furthermore, our wave geometry assessment mathematically rules out the viability of incident plane waves for high-localization ablation. Due to native exponential damping, a plane wave architecture generates an adverse thermal gradient where the skin surface receives nearly three times the heat dose delivered to the target focus, causing extensive collateral damage. In contrast, an active spherically focusing wave geometry introduces a geometric convergence factor ($\propto 1/r^2$) that dominates over exponential decay within a critical radius ($r_{\text{crit}} = 2x_0$), forcing a monotonic intensity increase along the propagation path. This active phase focusing explains the high clinical localization rates exceeding $90\%$, ensuring that surface tissues remain entirely uninjured.

Finally, by coupling this optimized acoustic field with a short-duration ($\tau = 1.0\text{ s}$) top-hat pulse, we solved the simplified non-diffusive Pennes bioheat equation. Integrating this ramp profile into the non-isothermal Arrhenius injury integral yielded a closed-form analytical expression for the sharp structural lesion boundary radius ($r_b = 0.75\, w_0$). Volumetric integration bounded tightly inside this cellular lesion perimeter tracks a stable spatial average temperature of $72.1^\circ\text{C}$ while point-wise values safely scale to a peak core maximum of $90^\circ\text{C}$, ensuring complete thermal destruction. Convolving the post-pulse relaxation field with a 2D free-space Green's function mathematically verified that the local temperature field at the boundary drops monotonically below the necrotic transition threshold ($60^\circ\text{C}$) within a sub-millisecond thermal relaxation window upon pulse termination. The analytical model confirms complete thermal and structural confinement, providing an explicit, hardware-independent framework to validate and optimize next-generation, high-tech non-invasive oncology devices.

{\begin{acknowledgments}
The author gratefully acknowledges support provided by the Gemini AI assistant (Google), including technical manuscript preparation, stylistic editing, and ensuring consistent United Kingdom English orthography.
\end{acknowledgments}

\section*{Declaration of Competing Interest}
The author declares that they have no known competing financial interests or personal relationships that could have appeared to influence the work reported in this paper.

\section*{Data Availability Statement}
All data and mathematical parameters required to evaluate the conclusions or reproduce the analytical solutions and numerical scaling trajectories are fully contained and presented within this manuscript. No external data sources or repositories were generated or utilized during this study.}

\bibliography{paper93}

\end{document}